\documentclass[prl,letterpaper,twocolumn,aps,epsf]{revtex4}

\usepackage{dcolumn,amsmath,amsfonts,graphicx,epsfig}

\def\comment#1{}

\newcommand{\beg}{\begin{eqnarray}}
\newcommand{\eee}{\end{eqnarray}}

\def\cm#1{}

\newcommand{\R}{{\mathbb{R}}}
\newcommand{\nBar}[1]{\overline{#1}} 
\newcommand{\la}{\label}

\newcommand{\en}{{ E}}

\newcommand{\f}{\frac}

\newcommand{\be}{\begin{equation}}
\newcommand{\ee}{\end{equation}}
\newcommand{\ba}{\begin{eqnarray}}
\newcommand{\ea}{\end{eqnarray}}
\newcommand{\beq}{\begin{equation}}
\newcommand{\eeq}{\end{equation}}
\newcommand{\bea}{\begin{eqnarray}}
\newcommand{\eea}{\end{eqnarray}}
\newcommand{\bastar}{\begin{eqnarray*}}
\newcommand{\eastar}{\end{eqnarray*}}

\newcommand{\An}{{\sf A}}

\begin{document}

\title{ 
Semi-Meissner state and neither type-I nor type-II superconductivity in multicomponent
systems.}
\author{
Egor Babaev${}^{1,2}$ and Martin Speight${}^3$}
\address{${}^1$Laboratory of Atomic and Solid State Physics, 
Cornell University  Ithaca, NY 14853-2501 USA\\
${}^2$Department of Physics, Norwegian University of Science and 
Technology, N-7491 Trondheim, Norway\\
${}^3$ School of Mathematics, University of Leeds, Leeds LS2 9JT, UK
}
\begin{abstract}
Traditionally, superconductors are categorized as type-I or type-II.\, 
Type-I superconductors support only Meissner and normal states, while
type-II superconductors form magnetic vortices in sufficiently strong
applied magnetic fields. Recently there has been much interest in
superconducting systems with several species of condensates, in fields
ranging from Condensed Matter to High Energy Physics.
 Here we show that
the type-I/type-II classification is insufficient for such multicomponent
superconductors.  We obtain solutions representing thermodynamically
stable vortices with properties falling outside the usual type-I/type-II
dichotomy, in that they have the following features: (i) Pippard
electrodynamics, (ii) interaction potential with long-range attractive and
short-range repulsive parts, (iii) for an $n$-quantum vortex, a non-monotonic ratio 
${E}(n)/n$ where ${E}(n)$ is the energy per unit length, (iv) energetic preference for
non-axisymmetric vortex states, ``vortex molecules". Consequently, these
superconductors exhibit
an emerging first order transition into a
``semi-Meissner" state, an inhomogeneous state comprising a mixture of
domains of two-component Meissner state and vortex clusters.
\end{abstract}
\maketitle

The formation of vortices in type-II superconductors subjected to a
magnetic field
is one of the  most remarkable phenomena occuring in condensed matter.
In all type-II superconductors (i.e.\ superconductors
where the GL parameter, which is the ratio of the magentic field 
penetration length
to the coherence lenth,  is $\kappa > 1/\sqrt{2}$ \cite{Abrikosov}) these 
vortices  share 
a set of properties:  an $n$-quantum vortex
is unstable with respect to decay into $n$ one-quantum vortices,
two vortices have purely repulsive interaction, and invasion of
vortices 
 under normal 
conditions is a second order phase transition
characterised by a critical value of the external magnetic field $H_{c1}$.

Vortices as solutions of the GL equations also exist formally in a type-I 
superconductor ($\kappa<1/\sqrt{2}$), 
but these vortices are thermodynamically unstable.
The special case $\kappa=1/\sqrt{2}$ is also very interesting
since, at this value of $\kappa$, vortices do not interact 
\cite{Bogomolny:1975de,Kr}. 
One should note,  however, that in real life systems
the situation is more complicated; experiments 
\cite{hubener} show that in certain materials with $\kappa 
\approx 1/\sqrt{2}$ there might exist a tiny attractive force between 
vortices 
at a certain distance. Such an interaction  was reproduced in a modified 
one-component GL model with additional terms in the regime 
$\kappa \approx 1/\sqrt{2}$ \cite{teor}.
We should also mention a long range van der Waals - type vortex attraction
in layered systems produced by thermal fluctuations or disorder 
\cite{attr2}.

Besides superconductivity, the vortex concept has a direct counterpart in 
High Energy Physics, called the Nielsen-Olesen string. Such strings have been
 considered in cosmology 
\cite{Hindmarsh:1994re} where they are expected 
to form during a symmetry breaking phase transition in the early universe. 
There also exists  a similar type-I/type-II 
division
of semilocal cosmic strings in the Higgs doublet model \cite{semilocal}.

Recently, multicomponent superconducting systems 
have attracted increasing interest in areas ranging
through metallic superconductors, hydrogen
in extreme conditions and color 
superconductivity in dense QCD \cite{frac,Nature,schmitt}. 
Below we consider a generic two-component superconductor (TCS), showing 
that it allows a novel 
type of thermodynamically stable 
vortices, whose electrodynamics is of Pippard type
with respect to one of the order parameters and
which have non-monotonic interaction energy
for a wide range of parameters as an intrinsic feature.
We will show that, as a result of this, such a TCS  
displays very unconventional magnetic properties which have no
counterparts in single-component systems,
and do not fall into the  standard type-I/type-II classification.

In the simplest case, the  
TCS (related to the two-Higgs  model \cite{tdlee}) 
 can be described by the following GL energy density:
\beg
&&
F = 
\frac{\hbar^2}{4m_1} \left| \left( \nabla +
i \f{2e}{\hbar c} {\bf A}\right) \Psi_1 \right|^2 + 
\frac{\hbar^2}{4m_2}  \left| \left( \nabla +
i \f{2e}{\hbar c} {\bf A}\right) \Psi_2 \right|^2 
 \nonumber \\
&&
+ { V} (|\Psi_{1,2}|^2)+ \eta [\Psi_1^*\Psi_2+\Psi_2^*\Psi_1] 
+ \frac{{\bf H}^2}{8\pi}
\la{act}
\eee
where $\Psi_\alpha = |\Psi_\alpha|e^{i \phi_\alpha}$ and $
{V} (|\Psi_{1,2}|^2)=\sum_{\alpha=1,2}-b_\alpha|\Psi_\alpha|^2+ 
\frac{c_\alpha}{2}|\Psi_\alpha|^4
$. 
In the model (\ref{act}),
vortices with phase winding in only one 
field have logarithmically divergent energy per
unit length if $\eta=0$ and linearly 
divergent if $\eta \ne 0$ \cite{frac}.
Here we do not consider the effect of thermal fluctuations, so
we can, without loss of generality, restrict attention to vortices with 
$\phi_1-\phi_2=const$ in the case $\eta=0$. These have finite-energy per 
unit length. Moving $\eta$ from zero 
will merely  change the core sizes of such vortices, so this is
a trivial extension. The same applies to other
possible additional potential terms in (\ref{act}).

The model (\ref{act}) possesses three characteristic length scales:
two coherence lengths $\xi_\alpha=\hbar/\sqrt{4m_\alpha b_\alpha}$
 and the magnetic field penetration 
length $\lambda=(c/\sqrt{8\pi} e)[ {|\nBar{\Psi}_1|^2}/{m_1}+
{|\nBar{\Psi}_2|^2}/{m_2}]^{-1/2}$, where  $|\nBar{\Psi}_\alpha|=
\sqrt{b_\alpha/c_\alpha}$.
It will also be convenient to define length parameters $\lambda_\alpha =
  c\sqrt{m_\alpha}/(\sqrt{8\pi}e{|\nBar{\Psi}_\alpha|})$, and 
thermodynamic critical magnetic fields for the
{\it individual} condensates,  
$H_{ct (\alpha)}=\Phi_0/(2\sqrt{2}\pi\xi_\alpha\lambda_\alpha)$. When 
the individual condensate $\Psi_1$ is of type-II,
we denote its  first and the second critical magnetic fields
as $H_{c1 (1)}=\Phi_0/(4 \pi \lambda_1^2)[\log (\lambda_1/\xi_1) + 0.08]$
and $H_{c2 (1)} = \Phi_0/(2 \pi \xi_1^2)$,
where $\Phi_0$ is the magnetic flux quantum.

Let us consider the magnetic properties of a TCS in several regimes.
In the simplest cases, when 
$\xi_1 \approx \xi_2 >> \lambda$ and when $\xi_1\approx \xi_2 << \lambda$, 
the magnetic properties of the TCS
parallel those of single-component type-I and type-II superconductors
correspondingly. 
However, we find that if one
of the condensates is of type-I while the other is of type-II,
the TCS in an external field has a much richer phase diagram.

In the case when $\xi_1 << \lambda_1$ and $\xi_2 >> \lambda_2$, and
$H_{ct(2)}$, the thermodynamic critical magnetic field  for the type-I 
condensate $\Psi_2$, is much higher than 
$H_{c2(1)}$ for the type-II condensate $\Psi_1$, the system  undergoes a
first order transition from a normal state 
immediately into a TCS
state. This is because, at fields higher than $H_{ct(2)}$,
the system does not allow nontrivial solutions of the 
linearized GL equation for $\Psi_1$, while, when 
the field is lowered below $H_{ct(2)}$, there appears 
a transition immediately into a
TCS state. In this state the magnetic 
field is screened in the bulk of the sample largely due to
surface current $\Psi_2$,  and nothing can preclude the appearance of the 
second condensate $\Psi_1$ in the bulk of the sample,
even if the applied field is much larger than $H_{c2(1)}$.  

Now consider 
the regime occuring when
$\lambda_1/\xi_1> 1/\sqrt{2};\lambda_2/\xi_2<1/\sqrt{2}; 
\xi_2 > \lambda$. 
Then a vortex solution should have an extended core 
associated with the condensate
$\Psi_2$ which exeeds the penetration length (the crucial question of whether
such vortices can be thermodynamically stable
will be answered below).
For a TCS, the 
 vortex energy consists of the  energies of the cores, the
kinetic energy of the Meissner current  
and the magnetic field energy. The magnetic field energy 
and the kinetic energy of the screening current are given by: 
$F_{m}=(1/8\pi )\int d^3 x {\bf H}^2 + (1/8\pi )\int d^3 x 
\lambda_{\rm eff}^2 ({ \rm curl} {\bf H})^2$.
Here we stress that, in the present case,  $|\Psi_2({\bf x})|^2$
varies slowly over the London penetration length $\lambda$. 
The magnetic field is screened at a distance 
from the core which is smaller than $\xi_2$. This means 
that only a depleted density of Cooper pairs of the condensate $\Psi_2$
participates in the screening of the magnetic field.
Thus, one cannot use the London penetration length $\lambda$,
but should introduce an effective penetration length 
$\lambda_{\rm eff} = [1/\lambda_1^2+1/{\tilde\lambda_2^2({\bf x})}]^{-1/2}$
where ${\tilde\lambda_2}({\bf x})=  c\sqrt{m_2}/( \sqrt{8\pi} e{|\Psi_2 
({\bf x})|}) > \lambda_2 = c\sqrt{m_2}/( \sqrt{8\pi} e{|\nBar{\Psi}_2|})$.
In the case  when $\lambda_1 << \xi_2$ and $\lambda_1$ is much smaller
than the Pippard length of $\Psi_2$
 we have 
$\lambda_{\rm eff}\approx \lambda_1$ which corresponds to the situation when
the magnetic field is screened mostly by condensate 
$\Psi_1$. 
On the other hand, in the case when $\xi_2 >> \lambda_1>>\lambda_2>>\xi_1$
the magnetic field can be screened at the
scale of the Pippard penetration length  
$\lambda=\lambda_2^P \approx (\lambda_2^2 \xi_2)^{1/3}$
of the condensate $\Psi_2$ (this expression is valid when 
$\lambda_2^P<<\lambda_1$).
In the above expression for $F_m$,
we cut off integrals at the distance $\xi_1$
from the center of the core in order to obtain an estimate 
of the vortex energy with logarithmic accuracy. Then the energy  per
unit length of a one-flux-quantum vortex is
\be
{E} \approx \left(\f{ \Phi_0}{4\pi\lambda_{\rm eff}}\right)^2 \log 
\f{\lambda_{\rm eff}}{\xi_1} + {\cal V}_{c1} +{\cal V}_{c2}
\label{ve}
\ee
where ${\cal V}_{c\alpha }$ are the energies of the cores per unit length
which are of order of magnitude of 
$[core \ size]\times[condensation \ energy] $.
The estimate of the core energy can also be expressed as: 
\be {\cal V}_{c\alpha } \approx \f{\pi \xi_\alpha^2 H_{ct(\alpha)}^2}{8\pi}
=  \f{ \Phi_0^2}{8\pi}\f{e^2|\Psi_\alpha|^2}{c^2 m_\alpha}
=\f{1}{4} \left(\f{ \Phi_0}{4\pi}\right)^2\f{1}{\lambda^2_\alpha}
\ee
Consequently, the energy of
two cores is ${\cal V}_{c1} +{\cal V}_{c2} \approx
 ({ \Phi_0}/{8\pi\lambda})^2$. Let us now assume that such vortices are thermodynamically stable 
(below
we demonstrate numerically that this is indeed the case). Then
a straightforward calculation of the  field $H_{c1}^0$ at which it becomes
energetically favorable to let a {\it single} vortex 
 into the superconductor gives:
\be
H_{c1}^0\approx \f{\Phi_0}{4\pi}\left[\f{1}{\lambda_{\rm eff}^2} \log 
\f{\lambda_{\rm eff}}{\xi_1} \right]
+\f{\Phi_0}{16\pi}\left[\f{1}{\lambda^2_1}+\f{1}{\lambda^2_2}\right]
\la{hc1}
\ee
However this characteristic field strength cannot be interpreted as the
 first critical magnetic field as in type-II
superconductors. In fact, in contrast to the type-II regime,
invasion of vortices
into a superconductor in the regime in question
should be accompanied
by a magnetization jump
and be of first order. This is because
the vortex { has an extended core at length scale 
$\xi_2>\lambda_{\rm eff}$ 
which gives rise 
to attraction between vortices 
$U \propto K_{0}(\sqrt{2}r/\xi_2)$ for $r>\lambda_{\rm eff}$,
where $K_{0}$ is the Bessel function.}
The attraction originates in winning in condensation
energy in the condensate $\Psi_2$ when outer cores overlap.
As we shall see below, the interaction potential has a repulsive
part at a shorter length scale of the order of $\lambda_{\rm eff}$.
Consequently, a lattice of vortices with a spacing 
determined by the minimum of the interaction potential is preferred over 
a system of widely separated  vortices.
 So  the energy 
of a system of $n$ vortices 
in this regime is minimized when vortices spontaneously form 
{\it lattice clusters}  with overlapping outer cores.
We should observe that $H_{c1}^0$ in the present situation is 
larger than the thermodynamic critical magnetic field of the 
type-I condensate, 
 $H_{ct (2)}=\Phi_0/(2\sqrt{2}\pi\xi_2 \lambda_2)$. 
We stress that this does not mean 
that at the field (\ref{hc1}) the condensate $\Psi_2$ is completely 
depleted, however. This is because, 
when  the applied  field  is close to  $H_{c1}^0$, 
the field is mostly screened by the supercurrent of the condensate $\Psi_1$,
which circulates along the  sample's edge. Thus the vortex system is dilute,
or, more precisely, the intervortex distance is
only determined by the effective attraction.
{\it So in contrast to the usual type-I / type-II behavior, the TCS in this regime displays a first order transition into an
inhomogeneous state consisting of clusters of vortices, where the order
parameter $\Psi_2$ is depleted due to the overlap of outer cores}. So 
superconductivity in these vortex  ``droplets" is dominated by the 
order parameter $\Psi_1$,
and is essentially a one-component superconductivity. Since the vortex 
density depends on the applied field, while the
intervortex distance is determined 
by the nonmonotonic interaction potential, there should be present,
besides these clusters of vortices, domains of  
two-component superconductivity in 
the vortexless Meissner state. We call
this the {\it semi-Meissner state}. 
The transition
into the semi-Meissner state may be viewed as a 
vortex matter analog of the condensation
of water droplets or a sublimation process in classical physics,
with the external field playing the role of ``pressure".
At a higher value of external field
the system will transition from
the semi-Meissner state to the regular Abrikosov lattice.
This transition might be rather complicated
because of the features of the vortex interaction 
potential. That is, the standard argument in favour
 of triangular lattice symmetry no longer
holds. Different lattice symmetries for a given
density of vortices are characterised by different
numbers of neighbors and different nearest-neighbour distances,
so one might 
construct a sequence of transitions between
lattices of different symmetries as
a function of external field strength.
Again an analogy between the external 
field strength and the role of pressure in classical physics 
might be invoked. 

{\it The key question 
regarding the existence of this transition and the semi-Meissner
state is whether vortices in this regime are stable
not only topologically but also  thermodynamicaly}.
Below we find an answer to this numerically.

For the purposes of numerical simulation of this system, 
it is convenient to rescale the fields. Let
$
\An=({2e}/{\hbar c}){\bf A}, \
\psi_\alpha=e/c\sqrt{(8\pi/m_\alpha)}\Psi_\alpha$.
Further, we introduce some new parameters
$
\mu_\alpha^2=(2\pi e^2 c^2/\hbar^2)m_\alpha^2c_\alpha,\qquad
u_\alpha^2=({4b_\alpha}/{\pi e^2c^2m_\alpha c_\alpha})$,
so that the effective potential becomes
$V_\alpha=\frac{\mu_\alpha^2}{8}(u_\alpha^2-|\psi_\alpha|^2)^2.$
Then the free energy density in the case $\eta=0$ is
\beq
\frac{16\pi e^2}{\hbar^2 c^2}F=
\frac{1}{2}|\nabla\times\An|^2+
\sum_{\alpha=1,2}\frac{1}{2}|(\nabla+i\An)\psi_\alpha|^2+
V_\alpha.
\eeq
In terms of the new parameters,
the 3 natural length scales are the inverse masses of the photon and the
two Higgs bosons, that is, $\lambda=(u_1^2+u_2^2)^{-\frac{1}{2}},\,
{\xi_\alpha}/{\sqrt{2}}=(\mu_\alpha u_\alpha)^{-1}$.
 This system supports radially symmetric solutions
\beq
\An=r^{-1}a(r)(-\sin\theta,\cos\theta),\qquad
\psi_\alpha=\sigma_\alpha(r)e^{ni\theta},
\eeq
where $a$ and $\sigma_\alpha$ are real and satisfy the boundary conditions
$
a(0)=\sigma_\alpha(0)=0$,
$a(\infty)=-n$,  $\sigma_\alpha(\infty)=u_\alpha$.
This ansatz reduces the field equations to an 
ODE system, 
\bea
a''-\frac{a'}{r}-(n+a)(\sigma_1^2+\sigma_2^2)&=&0
\nonumber 
\\
\label{ode}
\sigma_\alpha''+\frac{\sigma_\alpha'}{r}-(n+a)^2\frac{\sigma_\alpha}{r^2}+
\frac{\mu_\alpha^2}{2}\sigma_\alpha(u_\alpha^2-\sigma_\alpha^2)&=&0,
\eea
solutions of which may be found
by means of a shooting method similar to that used in \cite{spe}
(see remark \cite{remark1}).
Figures 1A--1C were generated using this scheme, employing
a 4th order Runge-Kutta method with variable 
$r$ step for the numerical integration,
and with $r_0=0.01$, $r_1=2$, $r_\infty=8$. The regime of most
interest is the  one where $\xi_1/\sqrt{2}<\lambda<\xi_2/\sqrt{2}$.
To explore the dynamical issues of interest,
 we would like to make the disparity between $\xi_1$ and $\xi_2$
as extreme as possible. There is a numerical limit to the disparity
of length scales we can achieve
(see remark \cite{remark1a}), but the phenomena of interest can be 
demonstrated within that limit.

Figure 1A shows the profile functions of a single two-component vortex in 
the neither type-I nor type-II regime, where the electrodynamics is of 
Pippard type in respect of component $\Psi_2$.
Note the disparity in healing lengths of the two condensate fields.
Figure 1B shows the energy of a single two-component vortex as a function
of $\xi_2$, with $\lambda$ and $\xi_1$, fixed, 
normalized by $\en_0$, the
energy of a {\em one-component} vortex with the same $\lambda$ and $\xi=
\xi_1$. Figure 1C shows the energy per vortex of $n=1$, $n=2$ and $n=3$
co-centered vortices at fixed $\xi_1$, $\xi_2$, $\lambda$, as a
function of $|\nBar{\Psi}_1|/|\nBar{\Psi}_2|$,
normalized by $\en_0$, the energy of a single one-component vortex with
the same
$\lambda$ and $\xi=\xi_1$. When $|\nBar{\Psi}_1|=|\nBar{\Psi}_2|$, the
$n=1$ vortex is favored, but as $|\nBar{\Psi}_1|/|\nBar{\Psi}_2|$ increases, first the
$n=2$ vortex then the $n=3$ vortex become energetically favored.
This is a very interesting property since such an effect has
no counterpart in type-II or type-I 
one-gap superconductors (N=1 Abelian Higgs model).
We should note that the $n=2,3$ co-centered solutions are unstable
with respect to formation of non-axisymmetric $n$-quantum vortex molecules, 
as we shall now see.

To compute the interaction energy of a vortex pair (shown
in Fig.\ 1D), one must go
beyond the radially symmetric ansatz and resort to a lattice
minimization method (see remark \cite{remark2}).
 Note that the intervortex force is attractive 
at long range but has a repulsive core, 
as predicted. The only stable static two-vortex we find
in this regime
has nonzero vortex separation and broken axial symmetry.
For an isolated one-quantum vortex, we find numerically that 
when $\xi_1 \ll \lambda \ll \xi_2$ and $\lambda_1$
is much smaller than the Pippard length of the condensate $\Psi_2$,
the vortex energy per unit length tends asymptotically to the energy of a 
vortex in a single condensate
$\Psi_1$ plus the core energy of the vortex of condensate $\Psi_2$.
Noting that the thermodynamical critical magnetic field
of $\Psi_1$ is proportional to $(\xi_1\lambda_1)^{-1}$,
while the core energy of the vortex in $\Psi_2$ is proportional to $\lambda_2^{-2}$,
this proves the thermodynamical stability of the 
neither type-I nor type-II vortices whose electrodynamics is
of Pippard type,
and the existence of the semi-Meissner state. 
\begin{figure}
\epsfxsize=3.4truein
\epsfbox[100 280 530 610]{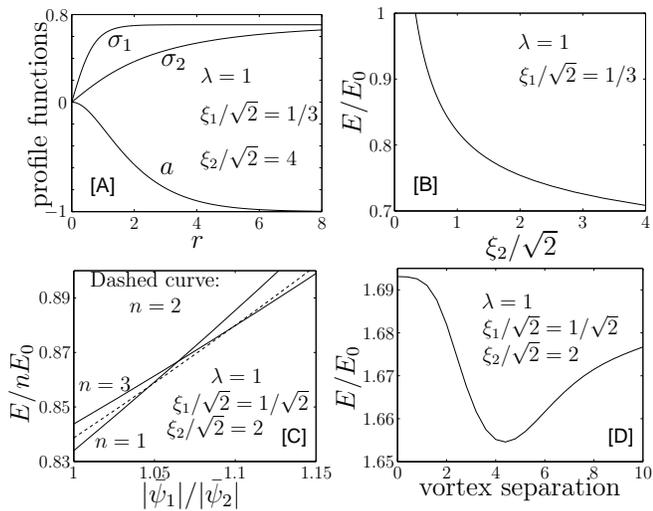}
\caption{{\bf A}: The numerical solution for the neither type-I nor type-II 
vortex. 
{\bf B}: The energy of a double-core vortex as a function of the disparity 
of the two coherence lengths. {\bf C}: The non-monotonic ratio
of energy per unit length to number of flux quanta 
for axisymmetric $n$-quantum vortices. {\bf D}: The non-monotonic interaction
potential for double-core vortices.}
\end{figure}

In conclusion, superconductivity 
in multicomponent systems has recently
attracted much interest in the physics
of condensed matter and beyond. Here 
we show
that,  in contrast to
ordinary superconductors, 
multicomponent systems 
allow for thermodynamically stable vortices
even in the Pippard regime. Moreover, their magnetic properties
 in a certain range
of parameters do not allow one to classify such
a superconductor as type-II or type-I.\, Rather,
it should legitimately be placed in a separate  class.
Such a novel type of superconductivity should be relevant for a variety of 
systems.
For example, it is well known that 
disparity of coherence lengths occurs naturally in two-band superconductors.
E.g. ${\rm MgB_2}$ and ${\rm Mg_{1-x}Al_{x}B_2}$, while type-II superconductors
not belonging to the regime we consider in this paper, have significant disparity
in coherence lengths \cite{mgb}.
Analogous situations might appear 
in mixtures of condensates with
different pairing symmetries. Another candidate
for this type of superconductivity is the projected
liquid metallic state of hydrogen
\cite{Nature} where this regime
is expected to be realized under certain conditions.
A similar situation
might also occur in the color superconducting state 
in quark matter \cite{schmitt}.  Certain features
of the considered state should be preserved
and might be experimentally accessed for vortex stacks  \cite{blatterRMP}  in
layered systems when one layer is
type-II and another is strongly type-I.

This work has been supported by STINT and Swedish Research Council,
 Research Council of Norway, Grant No.
157798/432 and National Science Foundation,
Grant DMR-0302347.


\begin{thebibliography}{99}
\bibitem{Abrikosov}
A.A. Abrikosov, Sov. Phys. JETP {\bf 5}, 1174 (1957)   

\bibitem{Bogomolny:1975de}
E.~B.~Bogomolny,
Sov.\ J.\ Nucl.\ Phys.\  {\bf 24}, 449 (1976)
see also 
J. Hove, S. Mo, and A. Sudb\o \  
Phys. Rev. B {\bf 66} 064524 (2002) 


\bibitem{Kr} L. Kramer Phys. Rev.B {\bf 3}, 3821 (1971).

\bibitem{hubener}
R.~P. H\"{u}bener { \it Magnetic Flux Structures in Superconductors}
Springer, Berlin (1979). 

\bibitem{teor}
{A.~E.} Jacobs, Phys. Rev. B { \bf 4},
3022 (1971) 
E.H. Brandt Phys. Stat. Sol.  {\bf 77}105 (1976);
I. Luk'yanchuk  Phys. Rev. B{\bf 63}, 174504 (2001),
F. Mohamed {\it et. al.}
Phys. Rev. B { \bf 65} 224504 (2002).

\bibitem{attr2}
G.\ Blatter and V.B.\ Geshkenbein,
Phys.\ Rev.\ Lett.\ \textbf{77}, 4958 (1996).
S.\ Mukherji and T.\ Nattermann,
{\it ibid} \textbf{79}, 139 (1997).


\bibitem{Hindmarsh:1994re}
M.~B.~Hindmarsh and T.~W.~Kibble,
Rept.\ Prog.\ Phys.\  {\bf 58}, 477 (1995)

\bibitem{semilocal}
A.~Achucarro and T.~Vachaspati,
Phys.\ Rept.\  {\bf 327}, 347 (2000)


\bibitem{frac}
E.~Babaev,
Phys.\ Rev.\ Lett.\  {\bf 89}, 067001 (2002). 
see also 
E.~Babaev, L.~D.~Faddeev and A.~J.~Niemi,
Phys.\ Rev.\ B {\bf 65}, 100512 (2002)

\bibitem{Nature}
E. Babaev, A. Sudb\o \  and N.W. Ashcroft Nature {\bf 431} 666 (2004),
E. Sm\o rgrav {\it et. al.} 
{\it Phys. Rev. Lett.} {\bf 94}, 096401 (2005)


\bibitem{schmitt}
K. Iida and G.  Baym.
{  Phys.\ Rev.\ {D}}  {\bf 66}, 014015 (2002);
A.~Schmitt,
nucl-th/0405076, page 115.


\bibitem{tdlee}
T.D. Lee, Phys.Rev. D {\bf 8 } 1226,  (1973)


%




\bibitem{spe}
J. M. Speight Phys.\ Rev.\ D {\bf 55} 3830 (1997)

\bibitem{remark1}
Since the system is singular at $r=0$ and $r=\infty$, we
shoot forwards from $r=r_0$, small, and backwards from $r=r_\infty$, large,
applying a matching condition at an intermediate radius $r_1$. 
The shooting parameters are the Taylor coefficients $s_i$ for the
small $r$ fields, and $q_i$,
the magnetic dipole/scalar monopole charges of the large $r$ fields
\cite{spe}: $
\hbox{small $r$:} \ \ a(r)=s_0 r^2+\cdots,\quad
\sigma_\alpha(r)=s_\alpha r^n+\cdots$ 
$
\hbox{large $r$:} \ \ a(r)=-n+q_0 rK_1(\sqrt{u_1^2+u_2^2}r)+\cdots,
\sigma_\alpha(r)=u_\alpha-q_\alpha K_0(\mu_\alpha u_\alpha r)+\cdots.$
The problem is thus reduced to zeroing an $\R^6$ valued function (the
mismatch at $r_1$ of $a,\sigma_1,\sigma_2,a',\sigma_1',\sigma_2'$)
on $\R^6$, which may be solved by a Newton-Raphson method. 
\bibitem{remark1a}
When $\xi_2$ is large, the field $\psi_2$ has a long range, so approaches
its vacuum value only slowly. We must thus choose $r_\infty$ to be large.
But then if $\xi_1$ is small, $u_1-\sigma_1(r_\infty)$ is tiny, and
large changes in $q_1$ are required to control the shots. Hence, the
Newton-Raphson scheme is close to singular, and the numerical method
becomes unstable. 
\bibitem{remark2}
The idea is to fix a value of
the vortex separation $\rho$, then minimize $\en=\int_{\R^2}F$ subject to the
constraint that $\psi_\alpha(-\rho/2,0)=\psi_\alpha(\rho/2,0)=0$, 
$\alpha=1,2$. We chose to place the system on a regular spatial
lattice (spacing $\delta x=\delta y=h$) with forward differences
replacing the partial derivatives occuring in $F$, then to solve the
constrained gradient flow equation for $\en$ using the Euler 
method with time 
step $\delta t$, until the configuration converged
to a minimum (up to numerical tolerance, that is, until the energy loss
per time step is less than $\en_{tol}$).
Figure 1D was produced using this scheme with $h=0.1$ and 
 $\delta t=0.01$ on a $201\times 101$
grid, with $\en_{tol}=10^{-5}$, which should be compared with a
typical two-vortex energy (in natural units) of $2\pi$. 
Again, 
practical considerations limit the length-scale disparity we can handle.
If $\xi_1$ is small, we need $h$ to be small so as to resolve the core
structure of the $\psi_1$ field. But then if $\xi_2$ is large, we need
a very big grid to ensure that the $\psi_2$ field experiences no 
significant boundary effects.



\bibitem{mgb}
 M. R. Eskildsen {\it et. al.} Phys. Rev. Lett. {\bf 89}, 187003 (2002);
S. Serventi {\it et. al.}  Phys. Rev. Lett. {\bf 93}, 217003 (2004). 
\bibitem{blatterRMP}
G. Blatter  {\it et. al.} Rev. Mod. Phys. {\bf 66}
1125 (1994)


\end{thebibliography}
\end{document}